\documentclass[12pt,fleqn,notitlepage]{article}
\usepackage{rotating}
\usepackage{booktabs}
\usepackage{bookmark}
\usepackage{graphicx}
\usepackage{caption}
\usepackage{hyperref}
\DeclareMathSymbol{,}{\mathord}{letters}{"3B}
\title{Surveying Residential Burglaries: A Case 
       Study of Local Crime Measurement}
\author{Robert Brame \& Michael G.\ Turner \\ 
        University of North Carolina at Charlotte \\ \\
        Raymond Paternoster \\ University of Maryland}
\date{January 2013}
\begin{document}
\maketitle
\begin{abstract} 
\noindent
We consider the problem of estimating the incidence of residential 
burglaries that occur over a well-defined period of time within 
the 10 most populous cities in North Carolina. Our analysis typifies
some of the general issues that arise in estimating and comparing
local crime rates over time and for different cities.
Typically, the only information we have about crime
incidence within any particular city is what that city's police 
department tells us, and the police can only count and describe 
the crimes that come to their attention. To address this,
our study combines 
information from police-based residential burglary counts and 
the National Crime Victimization Survey to obtain 
interval estimates of residential burglary incidence at the 
local level. We use those estimates as a basis for 
commenting on the fragility of between-city and over-time 
comparisons that often appear in both public discourse about 
crime patterns. 
\end{abstract}

\newpage
\section{Introduction}
It is difficult to count the incidence of crime at the local 
level.
Typically, the only systematically collected crime data are those
compiled by the local police department for submission to the
Federal Bureau of Investigation's (FBI) Uniform Crime Reporting
(UCR) program. These numbers are best suited for tracking the 
amount of crime that is reported to the police and how often 
these reported crimes are cleared by arrest or exceptional 
means (FBI, 2009). However, they often carry considerable 
weight in assessing how well a police department is performing, 
how safe a community is, and whether a police department needs 
more or different kinds of resources (Maltz, 1999:2). News 
media and law enforcement agencies routinely report levels and 
trends in crimes known to the police as ``crimes'' and 
``criminal behavior.'' In prepared testimony to the U.S.\ 
House of Representatives, Carbon (2012:16) writes that 
``[t]he UCR is the national `report card' on serious crime; 
what gets reported through the UCR is how we, collectively 
view crime in this country.''
Sometimes, these assessments veer into explicit crime rate 
rankings of cities (Rosenfeld and Lauritsen [2008] 
discusses the scope of this problem) -- a practice that has 
been condemned by the American Society of Criminology (2007). 
Criminologists sometimes use and compare point estimates 
of crime rates for different jurisdications at various 
levels of aggregation, and they report relationships 
between crime and various social and economic indicators 
as if there was no uncertainty in those estimates beyond 
sampling variation.\footnote{Such comparisons 
arise in fixed-effect regressions or difference-in-difference
estimators that attempt to statistically relate changes in
police-based crime statistics to changes in other variables 
that vary over time. A theme of this paper is that the
problem is greater than the usual concerns about random 
measurement error in the outcome variable. In fact, we believe
there is no basis for assuming random measurement error 
in this context.} In our view, there is no 
inherent problem with considering whether a crime rate is 
higher in one place than another at the same time or 
whether a crime rate is higher at one time than another 
in the same place. To be absolutely clear, the problem 
arises when ambiguities in the statistics undermine the 
validity of the comparison.

For example, a major obstacle to using police-based crime 
statistics to infer within-community changes in crime over time or 
between-community crime differences arises from the well-known 
fact that many crimes are not reported to the police (Baumer and 
Lauritsen, 2010; James and Council, 2008). Therefore, when 
crime rates vary across space or time, it is hard to know how 
much of that change is caused by shifts in real criminal 
behavior or changes in the reporting behavior of victims 
(Biderman and Reiss, 1967; Maltz, 1975; Eck and Riccio, 1979; 
Blumstein et al., 1991, 1992; also for a similar idea in
state SAT rankings see Wainer, 1986).

Consider a simple anecdote that illustrates our concerns. 
A recent newspaper article in the \emph{Charlotte Observer} 
reported that ``the number of crimes dropped 7.1 percent 
last year, a development that Charlotte Police Chief 
Rodney Monroe credited largely to officers keeping a close 
eye on potential criminals before they struck'' 
(Lyttle, 2012). The comparison expressed in this news 
coverage implicitly makes the strong and untestable assumption 
that the reporting behavior of victims stayed the same and all 
of the change in the number of crimes known to the police 
from one year to the next is due to changes in criminal 
behavior (Eck and Riccio, 1979; Blumstein et al., 1991;
Brier and Fienberg, 1980; Nelson, 1979; Skogan, 1974; 
Rosenfeld and Lauritsen, 2008; Bialik, 2010). 

Analytically, the same problems exist when criminologists 
try to explain variation in crime rates across different 
cities with identified explanatory variables like police 
patrol practices, dropout rates, home foreclosures, 
and unemployment; they also arise when researchers
try to measure and explain short-term changes in crime rates 
within the same jurisdiction. Whether the analysis involves
simple year-over-year percent change comparisons for different
cities or more complex statistical models for cross-sectional
and panel data sets, the analytical ambiguities are the same.

In fact, variation in crime reporting patterns injects 
considerable ambiguity into the interpretation of 
police-based crime statistics (Eck and Riccio, 1979; 
Blumstein et al., 1991; Levitt, 1998). Recent work by 
Baumer and Lauritsen (2010) -- building on a long series 
of detailed crime reporting statistics from the 
National Crime Survey (NCS) and its successor, the National 
Crime Victimization Survey (NCVS) -- makes the compelling 
point that there may be a causal relationship between 
the mobilization of the police and the likelihood that 
a community's citizens will report victimization 
experiences to the police. Police departments that 
cultivate strong working community partnerships may 
actually increase reporting of certain crimes simply 
because people believe the police will take useful 
actions when those crimes are reported:
\begin{quote}
Police notification rates are indicators of public confidence 
in the police and the legitimacy of the criminal justice 
system, and increasing police-public communication is a key 
goal of community-oriented policing strategies to reduce 
crime and the fear of crime (Baumer and Lauritsen, 2010:132).
\end{quote}
Even changes in the number of police in a particular area
may affect crime reporting practices of the citizenry
(Levitt, 1998).
Variation in reporting rates can create the 
illusion of a shift in crime even if real crime levels 
are perfectly stable (Eck and Riccio, 1979; Maltz, 1975). 
In fact, residential burglary reporting rates do exhibit 
year-to-year volatility. From 2010 to 2011, the rate 
at which residential burglary victimizations were 
reported to the police dropped from 59\% to 52\%
(Truman, 2011:10; Truman and Planty, 2012:9).
If these kinds of changes occur at 
the local level as well, they could easily explain a good 
deal of the variation we typically see from one year to the 
next in local, police-based robbery and burglary statistics.

In this paper, we conduct a case study of local level crime
measurement while trying to pay close attention to some 
important sources of ambiguity.
Specifically, our objective is to estimate the incidence 
of residential burglary for each of the 10 most populous 
cities in North Carolina in 2009, 2010, and 2011. The 
analysis is informed by 
data measuring the likelihood that residential burglary 
victimizations are reported to the police. We focus on 
residential burglaries in the 10 largest North Carolina 
cities because: (1) residential burglary is a clear, 
well-defined crime about which the public expresses 
considerable fear and concern (Blumstein and Rosenfeld, 
2008:18-20); (2) unlike most other states, North 
Carolina law enforcement agencies publicly report 
residential burglaries known to the police separately 
from non-residential burglary; (3) residential 
burglaries are household-level victimizations which 
correspond closely to the household-level structure 
of the NCVS (the NCVS does not measure reporting 
behaviors for commercial burglaries); and (4) 
conducting the analysis across cities and over time 
allows us to comment directly on the kinds of 
comparisons that are often conducted with police-based 
crime statistics. 

We are not the first to consider this 
issue (see, for example, Maltz, 1975; Eck and Riccio, 1979; 
Blumstein et al., 1991; Levitt, 1998; 
Lohr and Prasad, 2003; Westat, 2010;
Raghunathan et al., 2007). 
Nonetheless, the interval estimates or bounds we propose 
in this paper consider several key sources of 
uncertainty and stand in contrast to the excessively 
definitive point estimates that are often the subject of 
public discourse about crime.

\section{Legal Cynicism and Crime Statistics}
The first issue to address is why would we expect 
different rates of reporting crimes to the police in 
different jurisdictions (cities, counties, 
states)?\footnote{While our intention is not to offer an 
empirical test of the legal cynicism hypothesis or a 
definitive answer to the question regarding variablility 
in reporting to the police across jurisdictions, we do 
think that: (1) we owe the reader a plausible reason as 
to why this issue might matter, and (2) we think that at 
least part of the variation in reporting a crime to the 
police is due to differences in citizens' lack of 
confidence in the police -- part of what is thought of 
as legal cynicism.} Since most police work is reactive 
rather than proactive, questions about the relationship between
public perceptions of the police and the propensity 
of citizens to report crime victimizations to the
police loom large. If the public perceives the 
police as indifferent and unlikely to do anything to 
help, the likelihood of crimes being reported to the
police could be affected (Baumer and Lauritsen, 2010). 
Concerns that the police are unresponsive to the needs 
of the community can lead to a phenomenon called 
``legal cynicism.''  

Kirk and colleagues (Kirk and Matsuda 2011:444; Kirk and 
Papachristos 2011) have argued that legal cynicism is a 
``cultural frame in which the law and the agents of its 
enforcement are viewed as illegitimate, unresponsive, 
and ill equipped to ensure public safety.'' In addition, 
legal cynicism is understood to be ``an emergent property 
of neighborhoods in contrast to a property solely of 
individuals'' in part because it is formed not only in 
reaction to one's own personal experiences with legal 
actors and institutions but through interaction with 
others in the community (Kirk and Matsuda 2011:448). 
According to this view, culture, and legal cynicism 
as part of it, is not perceived as a set of values, 
goals, or ``things worth striving for'' (Merton 1968:187) 
but rather as a repertoire or toolkit to use in 
understanding the world (Swidler 1986).

There are two consequences that follow from the level 
of legal cynicism in a community. First, if legal 
institutions like the police are perceived as illegitimate 
then citizens are less willing to comply with laws with 
the result that there is going to be more actual crime. 
For example, Fagan and Tyler (2005) found that adolescents 
who perceived a lack of procedural justice among authorities
also exhibited higher levels of legal 
cynicism.\footnote{Fagan and Tyler's operational definition
of legal cynicism follows the original by Sampson and 
Bartusch (1998) which was conceptualized as a general 
sense of moral normlessness or a lack of respect for 
society's rules. Examples of items include ``laws are 
made to be broken'' and ``to make money, there are no 
right or wrong ways anymore, only easy ways and hard ways.'' 
In contrast, Kirk and colleagues focused  more narrowly on 
the legal dimension of cynicism, in which people perceive 
the law, and the police in particular, as illegitimate, 
unresponsive and ill equipped to ensure public safety 
(Kirk and Matsuda 2011:447). Examples of items include, 
``the police are not doing a good job in preventing crime 
in this neighborhood'' and ``the police are not able to 
maintain order on the streets and sidewalks in the neighborhood.''} 
Adolescents rated higher in legal cynicism (i.e., expressing
agreement with statements like ``laws are made to be broken'') 
were also higher in self-reported delinquency than those less cynical. 
In a survey of adult respondents, Reisig, Wolfe, and Holtfreter 
(2011) reported that self-reported criminal offending was 
significantly related to their measure of legal cynicism net 
of other controls including self-control. Finally, Kirk and 
Papachristos (2011) found that legal cynicism in Chicago 
neighborhoods explained why they had persistently high levels of 
homicide in spite of declines in both poverty and general 
violence. In addition to these studies of legal cynicism, 
there are numerous studies which have shown a link between 
measures of legitimacy of legal institutions such as the 
courts and police and a higher probability of violating 
the law (Paternoster et al.\ 1997; Tyler 2006; Papachristos, 
Meares, and Fagan 2011). 

A second consequence of legal cynicism -- of central concern 
in this paper -- is that citizens are 
not likely to cooperate with the police, including 
reporting a crime when it occurs. When citizens 
believe that the police are not likely to be responsive or 
will do little to help people like them, then we would 
expect more crimes to go unreported and offenders to go 
unarrested. The perception that it would do no good to 
cooperate with the police is an integral part of the 
cultural system described by Anderson (1999:323) as the 
``code of the street'': 
\begin{quote}
[t]he most public manifestation of this alienation is 
the code of the street, a kind of adaptation to a lost 
sense of security of the local inner-city neighborhood 
and, by extension, a profound lack of faith in the 
police and judicial system. 
\end{quote}
Several studies have found that community members -- 
both adults and juveniles -- are unlikely to cooperate 
with the police, including reporting crime and providing 
information, when law enforcement is seen as illegitimate 
(Sunshine and Tyler 2003; Tyler and Fagan 2008; 
Slocum et al.\ 2010). Kirk and Matsuda (2011) found a lower 
risk of arrest in neighborhoods with high levels of legal cynicism.

In sum, an increasing array of conceptual and empirical 
work has linked a perceived lack of responsiveness on the 
part of legal actors to both more crime and less reporting 
of crime. Communities characterized by high levels of legal 
cynicism or a sudden change in the level of legal cynicism 
(because of perceived mishandling of an event) may 
exhibit not only a higher level of crime but also a greater 
unwillingness of citizens to report a crime to the police. 
Given the reactive nature of most police work, there are
sound reasons for believing that citizens' lack 
of cooperation and faith in the police are reflected in 
a lower rate of official police-based crime statistics
though the actual rate may be higher.  
Although legal cynicism accounts for some of the 
variation in the rate at which citizens' report a crime to 
the police, other factors are also involved.
The point here is not to offer legal cynicism as the only 
hypothesis or even test this conjecture as a hypotheseis.
Rather, it is to provide some justification that there are 
credible \emph{a priori} reasons to believe there is systematic 
variation in the reporting of crimes across jurisdictions 
(and over time within the same jurisdiction) and that such 
variation is one source of the ambiguity in police statistics. 
In the following sections, we give formal expression to 
this ambiguity using an approach which brings the fragile nature 
of police-based crime statistics to center stage. 
  
\section{Police-Reported Residential Burglaries}
We begin our analysis by examining the number of 
residential burglaries reported by the police to the North 
Carolina State Bureau of Investigation's (SBI) 2010 
Uniform Crime Reports for the 10 most populous city-level
jurisdictions in North Carolina during the 2009-2011 period 
(State Bureau of Investigation, 2012). The SBI statistics
count both attempted and completed residential burglaries.
We verified that each of these 10 cities participated in the 
SBI's UCR program for each month of the 2009-2011 
calendar years. Table 1 identifies the 10 cities 
included in our study (column 1) along with the 
frequency of residential burglaries reported by 
each city's police department in 2009 (column 2), 
2010 (column 3), and 2011 (column 4). We denote residential 
burglaries reported by the police to the SBI-UCR
program as $b_p$.
\begin{table}
\begin{center}
\caption{Residential Burglaries Counted in the UCR $(b_p)$}
\begin{tabular}{lrrr}
\toprule
City & 2009 & 2010 & 2011 \\
\midrule
Asheville     &   545 & 457   &   555 \\
Cary          &   348 & 395   &   270 \\
Charlotte     & 7,766 & 7,305 & 6,352 \\
Durham        & 2,840 & 2,984 & 3,283 \\
Fayetteville  & 3,753 & 3,405 & 3,714 \\
Greensboro    & 3,766 & 3,487 & 3,279 \\
High Point    & 1,126 & 1,032 &   973 \\
Raleigh       & 2,488 & 2,442 & 2,364 \\
Wilmington    & 1,178 & 1,109 & 1,130 \\
Winston-Salem & 3,641 & 3,699 & 3,925 \\
\bottomrule
\end{tabular}
\end{center}
\end{table}
We formally attend to three ambiguities that arise
when these burglary numbers are presented as the 
actual number of burglaries committed (and the 
derivative burglary rate per 100,000 population)
in each jurisdiction: (1) the ``Hierarchy Rule'' for 
burglaries, (2) the population estimate used in estimating
the burglary rate; and (3) variation in the probability
that a residential burglary victim reports the incident
to the police.

The first issue we encounter in interpreting 
the burglary numbers in Table 1 is the UCR's 
Hierarchy Rule (Groves and Cork, 2008:173-175; 
Addington, 2007). 
The Hierarchy Rule mandates that any residential 
burglary reported to the police which co-occurs 
with an offense that ranks higher in the UCR 
hierarchy (aggravated assault, robbery, rape, or 
murder/non-negligent manslaughter) will not be 
counted as a residential burglary in the police 
statistics. Because of the Hierarchy Rule, we 
know the number of residential burglaries 
reported by the police to the UCR program, $b_p$, 
will generally be an undercount of the number of residential 
burglaries actually known to the police, which 
we denote as $b_k$. In order to estimate $b_k$ 
we need an estimate of the following fraction:
\begin{displaymath}
\theta = \frac{ \mbox{\# of Upgraded Residential Burglaries}}{b_p}
\end{displaymath}
so that $1+\theta$ provides us with an upward adjustment to
the police-reported counts to get an estimate of the number 
of burglaries known to the police:
\begin{displaymath}
b_k = b_p \times (1+\theta)
\end{displaymath}
Rantala (2000:7) measured crime classifications ignoring 
the Hierarchy Rule from the National Incident Based 
Reporting System (NIBRS) compared to crime classifications 
on the same criminal events using the traditional UCR approach.
This analysis was based on a detailed study of 
crime reports submitted by police departments 
covering 1,131 jurisdictions throughout the United 
States in the year 1996. According to the report, the 
participating police agencies reported 105,852 incidents 
of burglary (including both residential and commercial burglary) 
in the NIBR System. The number of burglaries that 
would have been counted if the UCR's Hierarchy Rule 
had been applied instead was 105,305 - which 
implies that the shrinkage in UCR-counted burglaries 
due to the Hierarchy Rule ($105,852-105,305 = 547$) 
is on the order of 
$\theta = \frac{547}{105,305} = 0.005$. 
An alternative analysis using more recent NIBRS 
data (2001) compiled by Addington (2007:239) suggests 
that $\theta$ attains an upper bound of 0.01. We
therefore assume that $\theta$ 
could be anywhere in the interval $[0.005,0.01]$. 
If our assumptions on the location and width of the 
$\theta$ interval are correct, the lower bound on $b_k$ will be:
\begin{displaymath}
\texttt{LB}(b_k) = b_p \times [1+\texttt{LB}(\theta)]
\end{displaymath}
while the upper bound on $b_k$ is:
\begin{displaymath}
\texttt{UB}(b_k) = b_p \times [1+\texttt{UB}(\theta)]
\end{displaymath}
These bounds show that the number of burglaries 
that appear in the police statistics slightly 
underestimate the number of burglaries actually known 
to the police. Table 2 presents the results of 
estimating bounds on $b_k$ in the 10 North Carolina cities.
\begin{table}
\begin{center}
\caption{Residential Burglaries Known to Police $(b_k)$}
\begin{tabular}{lrrrrrr}
\toprule
 & \multicolumn{2}{c}{2009} & 
   \multicolumn{2}{c}{2010} &
   \multicolumn{2}{c}{2011} \\
\cmidrule(r){2-3} \cmidrule(r){4-5} \cmidrule(r){6-7}
City & \multicolumn{1}{c}{\texttt{LB}($b_k$)} & 
       \multicolumn{1}{c}{\texttt{UB}($b_k$)} & 
       \multicolumn{1}{c}{\texttt{LB}($b_k$)} & 
       \multicolumn{1}{c}{\texttt{UB}($b_k$)} & 
       \multicolumn{1}{c}{\texttt{LB}($b_k$)} & 
       \multicolumn{1}{c}{\texttt{UB}($b_k$)} \\
\midrule
Asheville     &   548 &  550  &   459 &   462 & 558 & 561 \\
Cary          &   350 &  351  &   397 &   399 & 271 & 273 \\
Charlotte     & 7,805 & 7,844 & 7,342 & 7,378 & 6,384 & 6,416 \\
Durham        & 2,854 & 2,868 & 2,999 & 3,014 & 3,299 & 3,316 \\
Fayetteville  & 3,772 & 3,791 & 3,422 & 3,439 & 3,733 & 3,751 \\
Greensboro    & 3,785 & 3,804 & 3,504 & 3,522 & 3,295 & 3,312 \\
High Point    & 1,132 & 1,137 & 1,037 & 1,042 & 978 & 983 \\
Raleigh       & 2,500 & 2,513 & 2,454 & 2,466 & 2,376 & 2,388 \\
Wilmington    & 1,184 & 1,190 & 1,115 & 1,120 & 1,136 & 1,141 \\
Winston-Salem & 3,659 & 3,677 & 3,717 & 3,736 & 3,945 & 3,964 \\
\bottomrule
\end{tabular}
\end{center}
\end{table}
In each of the cities in our study, estimating $b_k$ 
based on $\theta$ and $b_p$ yielded only small increases
in the estimated number of residential burglaries. While the
practical effect of this adjustment is small, we
include it here in case new information ever appears
showing that our estimate of $\theta$ is too low. 

\section{Reporting Crimes to the Police}
The NCVS (and its predecessor, the NCS) is a nationally 
representative rotating panel household survey that has 
been continuously conducted by the Bureau of Justice 
Statistics and the U.S.\ Census Bureau since 1973. 
Households selected to participate in the survey 
remain in the sample for a 3.5 year period with 
interviews scheduled every 6 months for each household 
(although approximately $\frac{1}{6}$ of the 
households in the survey are interviewed each month of 
the year). For personal victimizations, the NCVS 
interviewer attempts to get information from each 
member of the household who is at least 12 years old. 
For property victimizations (including residential 
burglary), a designated person in the household 
answers all of the survey questions on behalf of the 
household.  

There is a large literature on the similarities and 
differences in crime trends measured by the UCR and 
the NCVS (Groves and Cork, 2008; Lynch and Addington, 
2007) and ways in which the UCR and the NCVS can be 
adjusted to better track each other (Rand and 
Rennison, 2002:50-51; Groves and Cork, 2008:74). 
What is unclear is how closely these trends track 
each other at \emph {local levels}. Since the NCVS 
can only be deployed for these types of comparisons 
in special cases and at specific times (Groves and 
Cork, 2008:73-74; Lauritsen and Schaum, 2005), 
there is no general way to answer this question -- 
it is an intrinsic source of uncertainty in our 
understanding of U.S.\ crime patterns.

In addition to asking survey respondents about 
victimization experiences, the NCVS interviewers ask 
those who experienced victimizations whether those 
incidents were reported to the police (Baumer and 
Lauritsen, 2010; Maltz, 1975).
The police reporting rates stratified by crime type 
are a standard table presented in NCVS reports going back to 
the original administrations of the NCS in 1973. In the
early years of the NCS, the reporting rates for residential
burglaries were usually estimated to be in the range of
45\% to 50\%. In recent years, however, the reporting
rates have mostly been in the range of 50\% to 59\%.
Generally speaking, then, reporting rates seem to have 
been trending upward over time. But even as recently as 
2006, the reporting rate fell below 50\% and in the most
recent year of data, the reporting rate was estimated
to be 52\%. Table 3 presents the residential burglary 
reporting rates (RR), estimated standard errors of those 
rates (se[RR]), and 95\% confidence limits for the 
2009-2011 period of our study.

\begin{table}
\begin{center}
\caption{Reporting Rates (2009-2011)}
\begin{tabular}{rrrrr}
\toprule
     &           &          & \multicolumn{2}{c}{95\% CL} \\
\cmidrule(r){4-5}
Year & RR      & se[RR] &    Lower & Upper \\
\midrule
2009 & 57.3 & 1.7 & 54.0 & 60.6 \\
2010 & 58.8 & 1.9 & 55.1 & 62.5 \\
2011 & 52.0 & 1.8 & 48.5 & 55.5 \\
\bottomrule
\end{tabular}
\end{center}
\end{table}

The reporting rate is an approximation to a critical 
parameter for our study -- the probability that a 
residential burglary victim reports that victimization 
to the police -- which we denote as $p_r$. For the
three years, we study closely in this paper, we ignore
the problem of respondents who either did not know or 
did not say whether a burglary victimization was reported
to the police. We assume that each city's $p_r$ lies 
within the 95\% confidence interval of the percent 
of residential burglaries reported to the police in 
each of the three years, 2009-2011.
 
\section{Bounding the Number of Burglaries}
When researchers, police chiefs, newspaper reporters, and 
booksellers report over-time changes and area 
differences in crimes known to the police as actual 
crime levels, they are making the strong assumption
that the probability of a crime victim reporting a 
victimization to the police is constant either 
over time or space (or perhaps both).  Let us 
return briefly to the \emph{Charlotte Observer} 
example cited earlier. That article reported that 
``the number of crimes dropped by 7.1\% last year.'' 
The only way this statement could be correct is if 
the reporting probability, $p_r$, was exactly the 
same from one year to the next. Since the 
reporting rates in Charlotte for the range of 
different crimes included in the \emph{Observer}'s 
article are not well understood, there is no 
justification for the certitude that $p_r$ 
stayed exactly the same from one year to the next. 
Furthermore, when researchers estimate difference-in-difference
regression models or pooled cross-sectional time-series
models to study the effects of this or that intervention,
social, or economic trend on crime patterns, the complexity 
of the comparisons and ambiguities in identification 
propagate and the analytical difficulties remain.

We think it is preferable to develop a plausible range 
of values for $p_r$ for those crimes. The range should 
have two key features: (1) it includes the true value 
of $p_r$; and (2) it expresses the uncertainty we have 
about the true value of $p_r$. Our approach assumes 
that the reporting rate for each individual city lies 
within the 95\% confidence interval of the reporting
rate estimated by the NCVS in each year. A useful 
consequence of our approach is that it will not be possible 
to express the incidence of residential burglary in terms 
of a single number. 
Instead, we adopt the perspective of Manski 
(1995, 2003) and argue that a range of estimates 
based on a weak but credible assumption is 
preferable to a fragile point estimate based on 
a strong and untestable assumption. 
Even if some would take issue with the precise 
boundaries of our interval, we still believe it to be 
far more credible than a constant $p_r$ assumption 
across cities or over time within the same city. 
And a key feature of our approach is that 
it is easy to use the information we present to calculate 
estimates based on different boundaries if there are good 
reasons for doing so.

We now turn to the task of estimating the 
actual number of residential burglaries in the 
10 North Carolina cities in 2009-2011. We 
refer to this estimand as $b_a$ where the 
subscript, $a$, means actual. What we have 
obtained so far are bounds on $b_k$ which is 
the number of residential burglaries known to 
the police, allowing for uncertainty due to the
UCR Hierarchy Rule. So, we need a way to move 
from the bounds on $b_k$ to bounds on $b_a$. 
In a world where $p_r$ and $b_k$ are 
known with certainty, we would estimate
\begin{displaymath}
b_a = \frac{b_k}{p_r}
\end{displaymath}
as discussed in Eck and Riccio (1979:298).
For example, if a community's police department 
recorded $b_k = 1,000$ residential burglaries in
2009 and the probability that a residential burglary 
victim reports the incident to the police is 
0.573 (the 2009 NCVS-estimated reporting rate) 
then it follows that the number of actual 
residential burglaries in 2009 would be:
\begin{displaymath}
b_a = \frac{b_k}{p_r} = \frac{1,000}{0.573} = 1,745
\end{displaymath}
Since neither $p_r$ nor $b_k$ is known with certainty, 
our approach is to place bounds on $b_a$. If we have 
$\theta \in [0.005,0.01]$ uncertainty due to the UCR 
Hierarchy Rule, our best estimate of the lower bound 
of $b_a$ is to divide the lower bound of $b_k$ by the 
upper bound of $p_r$ which we assume to be 0.606. 
This yields a lower bound on the estimated number of 
residential burglaries:
\begin{displaymath}
\texttt{LB}(b_a) = 
\frac{\texttt{LB}(b_k)}{\texttt{UB}(p_r)} = 
\frac{1,005}{0.606} = 1,658
\end{displaymath}
Conversely, the upper bound estimate is attained when:
\begin{displaymath}
\texttt{UB}(b_a) = 
\frac{\texttt{UB}(b_k)}{\texttt{LB}(p_r)} = 
\frac{1,010}{0.540} = 1,870
\end{displaymath}
So, in this instance, one could adopt the certitude 
that $p_r$ is exactly equal to 0.573, which would yield
a point-estimated number of 1,745 residential 
burglaries. By contrast, our approach is to make 
the following argument: (1) based on the available
evidence, if we had drawn repeated NCVS samples, we 
infer that 95\% of the samples would have produced
an estimated reporting rate between 54.0\% and 60.6\%;
(2) we do not have local-area estimates of the reporting
rates for individual cities; (3) we assume that the
individual cities are within sampling error of the 
national estimates; (4) if our boundaries 
on $p_r$ include the true value of $p_r$ and our 
assumptions about the Hierarchy Rule are correct 
then the number of actual burglaries lies 
between 1,658 and 1,870; and (5) our answer is 
less precise than a point estimate but our statement 
summarizing the results is far more likely to be 
correct and does a better job of transmitting both 
what we know and what we don't know to our audience. 

Applying this logic to the residential burglary 
data in the 10 North Carolina cities in 2009-2011, 
we estimate the bounds on the actual number of 
residential burglaries (Table 4), $b_a$.
The results in Table 4 highlight some key features of
our approach. Let's consider Charlotte as an example.
From Table 1, in 2010, we see that the Charlotte-Mecklenburg
Police Department reported 7,305 residential burglaries
while that number dropped to 6,352 burglaries in 2011 (a 13\%
year-over-year decline).
Yet Table 4 shows that the actual number of residential
burglaries in Charlotte in 2010 was between 11,742 and 13,396
while in 2011 it was between 11,496 and 13,221. Based on this
evidence it is not possible to definitively say whether the 
number of residential burglaries stayed the same, increased,
or decreased from 2010 to 2011. A critical source of ambiguity
in this comparison is that the 95\% confidence limits for the
reporting rate in 2010 and 2011 are $[0.551,0.625]$ and 
$[0.485,0.555]$, respectively. So, it is not possible to tell
whether the decline in Charlotte burglaries reported by the 
police is due to real changes in burglary or changes in the 
reporting rate from one year to the next (or both).

Asheville serves as a counterexample. In 2010, the Asheville Police
Department reported 457 residential burglaries while in 2011,
the number increased to 555 (a 21.4\% increase). Our analysis
estimates that the actual number of residential burglaries
increased from the range of $[735,838]$ in 2010 to the range
of $[1004,1156]$ in 2011. This means that the sign of the change
in Asheville is identified (positive) even with the uncertainty 
that we have allowed for the reporting rates.
The bottom line of Table 4 is that sometimes relatively
strong conclusions are warranted -- and sometimes they are not.

\begin{table}
\small
\begin{center}
\caption{Bounds on the Actual Number of Burglaries $(b_a)$}
\begin{tabular}{lrrrrrr}
\toprule
 &  \multicolumn{2}{c}{2009} 
 & \multicolumn{2}{c}{2010} 
 & \multicolumn{2}{c}{2011} \\
\cmidrule(r){2-3} \cmidrule(r){4-5} \cmidrule(r){6-7}
City & \multicolumn{1}{c}{\texttt{LB}($b_a$)} & 
       \multicolumn{1}{c}{\texttt{UB}($b_a$)} & 
       \multicolumn{1}{c}{\texttt{LB}($b_a$)} & 
       \multicolumn{1}{c}{\texttt{UB}($b_a$)} &
       \multicolumn{1}{c}{\texttt{LB}($b_a$)} & 
       \multicolumn{1}{c}{\texttt{UB}($b_a$)} \\
\midrule
Asheville     &    903  &  1,020 &    735 &    838 &  1,004 &  1,156 \\
Cary          &    577  &    651 &    635 &    724 &    489 &    563 \\
Charlotte     & 12,872  & 14,534 & 11,742 & 13,396 & 11,496 & 13,236 \\
Durham        &  4,707  &  5,315 &  4,796 &  5,472 &  5,942 &  6,841 \\
Fayetteville  &  6,221  &  7,024 &  5,473 &  6,244 &  6,722 &  7,739 \\
Greensboro    &  6,242  &  7,048 &  5,605 &  6,395 &  5,935 &  6,832 \\
High Point    &  1,866  &  2,107 &  1,659 &  1,893 &  1,761 &  2,027 \\
Raleigh       &  4,124  &  4,656 &  3,925 &  4,478 &  4,279 &  4,926 \\
Wilmington    &  1,953  &  2,205 &  1,783 &  2,034 &  2,045 &  2,355 \\
Winston-Salem &  6,035  &  6,814 &  5,946 &  6,783 &  7,104 &  8,178 \\
\bottomrule
\end{tabular}
\end{center}
\end{table}

\section{Population and Household Estimates}
In the UCR, burglaries are defined in terms of the 
number of incidents per 100,000 population. The NCVS,
on the other hand defines the burglary rate in terms
of the number of incidents per 1,000 households.
Regardless of which approach one adopts,
it is necessary -- at least as a starting point -- 
to have reasonable estimates of the number of persons 
living within a police department's jurisdiction 
(Gibbs and Erickson, 1976).
Both the SBI's and the FBI's UCR programs publish 
these estimates. We refer to the SBI estimate as $n_s$ 
while the FBI's estimate is denoted as 
$n_f$.\footnote{The Greensboro Police Department's 2011
data were published in the North Carolina State Bureau of
Investigation's Crime Reporting Program but not in the 
FBI's Uniform Crime Reporting Program. Consequently, we
do not have a FBI estimate of the size of the population 
in the jurisdiction of the Greensboro Police Department 
for the year 2011.}
Table 5 presents a summary of the two sets of population 
estimates for each of the cities in each year of our study.

\begin{table}
\footnotesize
\begin{center}
\caption{Population Estimates}
\begin{tabular}{lrrrrrr}
\toprule
 &    \multicolumn{2}{c}{2009} & 
      \multicolumn{2}{c}{2010} &
      \multicolumn{2}{c}{2011} \\
\cmidrule(r){2-3} \cmidrule(r){4-5} \cmidrule(r){6-7}
City & 
       \multicolumn{1}{c}{$n_s$} & 
       \multicolumn{1}{c}{$n_f$} & 
       \multicolumn{1}{c}{$n_s$} & 
       \multicolumn{1}{c}{$n_f$} &
       \multicolumn{1}{c}{$n_s$} & 
       \multicolumn{1}{c}{$n_f$} \\
\midrule
Asheville     &  78,267 &  74,923 &  78,804 &  83,393 &  82,846 &  84,450 \\ 
Cary          & 141,269 & 133,757 & 147,282 & 135,234 & 136,203 & 136,949 \\ 
Charlotte     & 738,768 & 777,708 & 752,799 & 779,541 & 776,787 & 789,478 \\
Durham        & 221,675 & 227,492 & 227,524 & 228,330 & 222,978 & 231,225 \\ 
Fayetteville  & 205,285 & 173,995 & 205,555 & 200,564 & 206,132 & 203,107 \\ 
Greensboro    & 257,581 & 253,191 & 261,519 & 269,666 & 263,279 & Missing \\ 
High Point    & 100,648 & 103,675 & 102,216 & 104,371 & 104,788 & 105,695 \\ 
Raleigh       & 367,514 & 406,005 & 373,100 & 403,892 & 395,716 & 409,014 \\
Wilmington    &  99,485 & 101,438 &  99,911 & 106,476 & 104,422 & 107,826 \\ 
Winston-Salem & 222,574 & 230,978 & 229,338 & 229,617 & 224,566 & 232,529 \\ 
\bottomrule
\end{tabular}
\end{center}
\end{table}

In looking at the estimates in Table 5 we are struck by how 
similar they are in some cases (for example, Winston-Salem and
Durham in 2010) and how different they are in 
others (for example, Fayetteville in 2009, and Charlotte and
Raleigh in 2009-10). 
We do not take a position on the comparative validity 
of the two population estimates, but the fact that they are 
sometimes not close to equal 
adds yet another layer of uncertainty to our interpretation 
of the crime rate. That there would be variation in 
population estimates when those estimates are compiled 
by independent agencies is unsurprising. Summarizing 
the size of the population or the number of households
in a particular police department's jurisdiction in 
a year's time with a single number must, on its face, 
be an approximation (Gibbs and Erickson, 1976).
It is also interesting that textbook discussions of 
the crime rate often point to ambiguities in counting 
the number of crimes but the 
ambiguities of counting the target population for any 
particular crime rate are a less prominent consideration 
(Lab et al.\ 2008:4-5). 

A second issue is the calibration of household
counts within a particular city. One could take the 
position of the NCVS, that residential burglary is a 
crime against an entire household and that residential 
burglaries are best expressed in terms of the risk per 
household rather than an individual per-person (or a 
scaled-up divisor such as 1,000 persons or 100,000 persons).
A question that needs to be considered in any specific 
analysis is whether the conclusions we draw depend on 
the scaling unit. It is useful to consider the polar 
case where the scaling unit would not create any 
ambiguity. If we are able to assume that the number 
of persons per household is constant over time within 
the same city and across cities, then the choice 
between scaling units (persons or households) is 
arbitrary. On the other hand, if the number of persons 
per household varies over time within a jurisdiction or 
between jurisdictions, then our rate estimators should 
accomodate this variability.

Table 6 relies on the data in Table 5 combined with information
from the U.S.\ Census Bureau's (2012) measure of the average number 
of persons per household (\texttt{pph}) in each of the 10 North 
Carolina cities over the period 2006-2010 to produce state and 
federal estimates of the number of households in each city 
($h_s$ and $h_f$, respectively). 

\begin{table}
\footnotesize
\begin{center}
\caption{Household Estimates}
\begin{tabular}{lrrrrrrr}
\toprule
 &  & \multicolumn{2}{c}{2009} & 
      \multicolumn{2}{c}{2010} &
      \multicolumn{2}{c}{2011} \\
\cmidrule(r){3-4} \cmidrule(r){5-6} \cmidrule(r){7-8}
City & \multicolumn{1}{c}{\texttt{pph}} &
       \multicolumn{1}{c}{$h_s$} & 
       \multicolumn{1}{c}{$h_f$} & 
       \multicolumn{1}{c}{$h_s$} & 
       \multicolumn{1}{c}{$h_f$} &
       \multicolumn{1}{c}{$h_s$} & 
       \multicolumn{1}{c}{$h_f$} \\ 
\midrule
Asheville     & 2.13 &  36,745 &  35,175 &  36,997 &  39,152 & 38,895 & 39,648 \\
Cary          & 2.68 &  52,712 &  49,909 &  54,956 &  50,460 & 50,822 & 51,100 \\
Charlotte     & 2.46 & 300,312 & 316,141 & 306,016 & 316,887 & 315,767 & 320,926 \\
Durham        & 2.30 &  96,380 &  98,910 &  98,923 &  99,274 & 96,947 & 100,533 \\
Fayetteville  & 2.48 &  82,776 &  70,159 &  82,885 &  80,873 & 83,118 & 81,898 \\
Greensboro    & 2.34 & 110,077 & 108,201 & 111,760 & 115,242 & 112,512 & Missing \\
High Point    & 2.51 &  40,099 &  41,305 &  40,724 &  41,582 &  41,748 & 42,110 \\ 
Raleigh       & 2.35 & 156,389 & 172,768 & 158,766 & 171,869 & 168,390 & 174,049 \\
Wilmington    & 2.19 &  45,427 &  46,319 &  45,621 &  48,619 &  47,681 & 49,236 \\
Winston-Salem & 2.42 &  91,973 &  95,445 &  94,768 &  94,883 & 92,796 & 96,086 \\ 
\bottomrule
\end{tabular}
\end{center}
\end{table}

A prominent aspect of the evidence in Table 6 is that there is real
variation in the number of persons per household in large North 
Carolina cities. On average, Asheville and Wilmington have smaller 
household sizes ($< 2.2$ persons per household) while Cary and High 
Point have the highest density households ($> 2.5$ persons per household).
It is possible, then, that two cities could have an identical rate of
residential burglary when that rate is expressed in terms of population
size but have different residential burglary rates when expressed in
terms of the number of households in the city. 

\section{Estimating Residential Burglary Rates}
We now consider how much our inferences about residential burglary 
rates depend upon the issues we have considered in this paper. We
don't expect much sensitivity to the uncertainty of the Hierarchy
Rule since the adjustments are small. It is less clear how sensitive
our results will be to uncertainty about the size of the population 
(including whether we scale by the number of persons or the number 
of households) and uncertainty about the fraction of residential 
burglaries reported to the police.

In order to estimate the actual rate of residential burglaries 
per 100,000 persons ($r_a$), we define its lower bound as:
\begin{displaymath}
\texttt{LB}(r_a) = 
\frac{\texttt{LB}(b_a)}
{\max(n_s,n_f)} \times 100,000
\end{displaymath}
while the upper bound is:
\begin{displaymath}
\texttt{UB}(r_a) = 
\frac{\texttt{UB}(b_a)}
{\min(n_s,n_f)} \times 100,000
\end{displaymath}
This interval estimate identifies the outer limits 
of what is possible in terms of the residential 
burglary rate per 100,000 persons assuming that
$n_s$ and $n_f$ form the proper bounds on the size
of the population, that each city's reporting rate
falls within the 95\% confidence interval of the
NCVS-estimated reporting rate, and that our adjustments
for the FBI's Hierarchy Rule are accurate. We combine
these estimates with the standard residential burglary
rates per 100,000 population based on the information
from Tables 1 and 5 to produce the point and interval
estimates in Figure 1.

\begin{figure}[!htbp]
\centering
\includegraphics[scale=0.55]{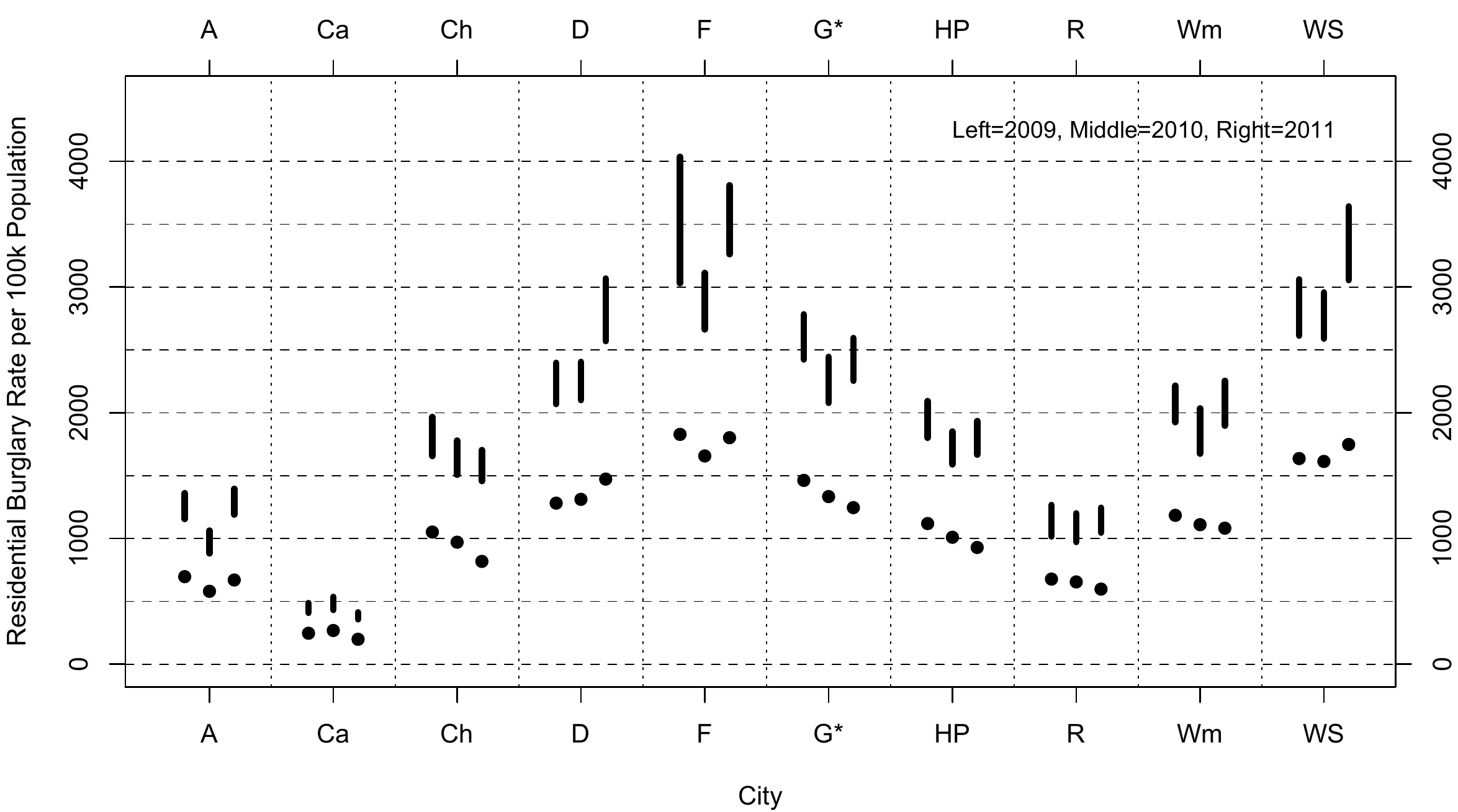}
\caption{Residential Burglary Rate (Population) Estimates}
\end{figure}
\begin{figure}[!htbp]
\centering
\includegraphics[scale=0.55]{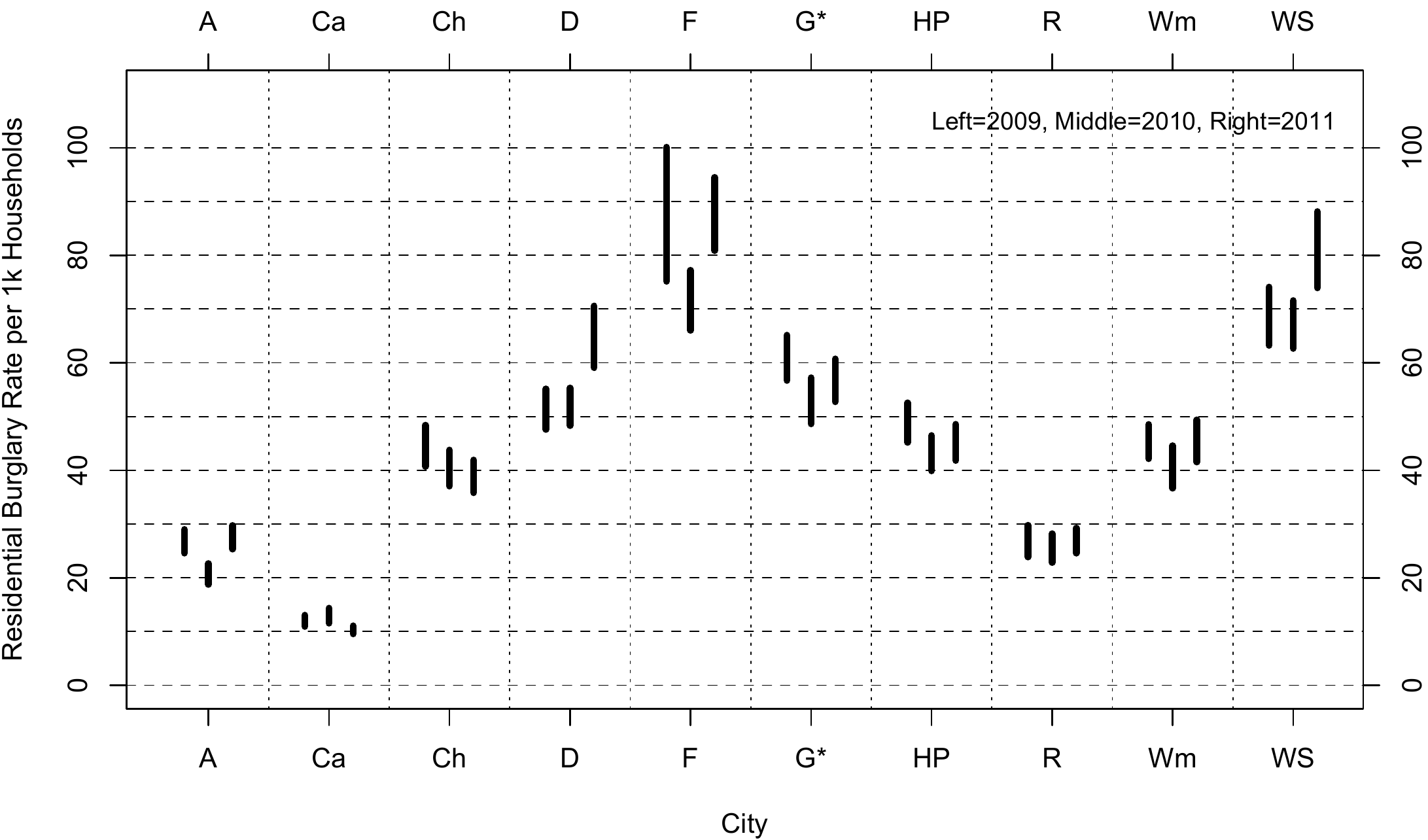}
\caption{Residential Burglary Rate (Household) Estimates}
\end{figure}

For each of the 10 cities in Figure 1 there are 2 sets 
of estimates: (1) the upper and lower bound (interval)
estimates of the actual residential burglary rate ($r_a$) for 
2009-2011; and (2) the ``standard'' (point) estimates of the 
residential burglary rate per 100,000 population (based 
on Tables 1 and 5) for 2009-2011.\footnote{We remind the reader
that Greensboro's analysis for 2011 is incomplete since the FBI
did not publish population estimates for that city in that year.} 
Comparisons between the point estimates over time within the 
same city implicitly assume that the reporting rate, $p_r$ is 
the same across the years while comparisons between point 
estimates of different cities implicitly assume that the 
reporting rate is constant between the cities being compared 
(at the time they are compared). While these assumptions 
seem implausibly strong, they are commonly invoked for both 
journalistic and research purposes.

We also consider the impact of adjusting for the number of 
households instead of the number of people and then placing 
residential burglaries on a scale per 1,000 households in
each city during each year (Figure 2). This analysis relies 
on the information presented in Tables 4 and 6. To estimate the
lower bound on the residential burglary rate per 1,000 households
we obtain:
\begin{displaymath}
\texttt{LB}(r_a) = 
\frac{\texttt{LB}(b_a)}
{\max(h_s,h_f)} \times 1,000
\end{displaymath}
and the upper bound is given by:
\begin{displaymath}
\texttt{UB}(r_a) = 
\frac{\texttt{UB}(b_a)}
{\min(h_s,h_f)} \times 1,000
\end{displaymath}
Broadly speaking, the two sets of rate comparisons in 
Figures 1 and 2 seem to tell similar stories. From this
analysis, our major conclusion is that the major source
of uncertainty in estimating residential burglary rates
in these 10 North Carolina cities over the 2009-2011
time frame is the reporting rate, and to a lesser extent,
the size of the population. 

It is worth considering a couple of example implications
of our results. Suppose we set out to compare the burglary
rates between Charlotte and Wilmington in 2009. Using the 
standard approach for comparing the two cities, we would 
find that Charlotte had a residential burglary rate of 1,051
per 100,000 persons while Wilmington's rate is 1,184 (nearly
13\% higher). Some might use this evidence to say that Wilmington
had a higher residential burglary rate than Charlotte in 2009.
But upon further analysis, we find that the actual rate of
residential burglaries per 100,000 population could plausibly
lie in the $[1656,1967]$ interval in Charlotte and the $[1925,2216]$
interval in Wilmington. Since these intervals overlap (see Figure
1), the sign of the difference between Charlotte and Wilmington
is not identified. 

This lack of identifiability becomes even more prominent
when we focus on household incidence of burglary. For Charlotte,
the bounds on the residential burglary rate per 1,000 households
in 2009 are $[40.72,48.40]$; for Wilmington, the bounds are
$[42.16,48.53]$. This is a marked increase in the degree of
overlap between the Charlotte and Wilmington interval estimates.
We can attribute most of this difference to the fact that 
Charlotte households had an average of 2.46 persons in 2006-2010
while Wilmington households were smaller on average 
(2.19 persons) over the same time period.
In short, to speak about a clear difference in these
two cities is an example of unwarranted certitude. It is possible
that Charlotte's burglary rate is higher, lower, or the same as
Wilmington's. Considering plausible sources of uncertainty in 
the comparison, the data are simply not strong enough to tell us. 

A comparison of Charlotte and Raleigh -- on the other hand -- 
leads us to a stronger set of conclusions. In 2011, for 
example, Charlotte's estimated residential burglary rate 
per 100,000 population -- based exclusively on the information 
in Tables 1 and 5 -- was 818 while Raleigh's rate was 597 
(a $\frac{818-597}{818} \times 100 \approx 27\%$ difference). 
We conclude that the difference between the burglary rates in Charlotte 
and Raleigh cannot be explained by the uncertainties considered in
this paper. The residential burglary rate interval for Charlotte
in 2011 is $[1456,1704]$ while the interval for Raleigh in the
same year is $[1046,1245]$. As Figure 2 shows, there is a 
similar pattern for burglary incidence scaled by the number of
households.  Since these intervals do not overlap, it
seems credible to argue that Charlotte's rate is higher than
Raleigh's rate in 2011.\footnote{We leave aside the question of 
\emph{why} Charlotte's rate is higher than Raleigh's rate. There
are a large number of possibilities. What we have been able to
establish with this analysis is that the difference between the
two cities cannot be explained by uncertainty in the population
size and sampling variation in the NCVS reporting rate. Thus, the
explanation(s) for the difference between the cities lies elsewhere.}

Figures 1 and 2 are also helpful for displaying the over-time
change within cities. Using the standard crime rate estimator, 
we can see that the point estimates of Charlotte's residential 
burglary rate reveal what appears to be a meaningful decline 
from 2009 (1,051) to 2011 (818) 
(a drop of about $\frac{818-1,051}{1,051} \times 100 \approx 22.2\%$). 
The problem is that there is some evidence that reporting rates could
have also changed a good deal over the same time period. What this
means is that the burglary rate interval for Charlotte (accounting
for the Hierarchy Rule, reporting rate uncertainty, and population 
size uncertainty) is $[1655,1967]$ in 2009 and $[1456,1704]$ in
2011. Since these intervals overlap we cannot discern whether the
Charlotte burglary rate increased, decreased, or stayed the same
over this time period. A counterexample is provided by Durham. In
2009, the standard burglary rate estimate was 1,281; in 2011 that 
rate estimate increased to 1,472 -- an increase of 
$\frac{1,472-1,281}{1,281} \times 100 \approx 14.9\%$. Our interval
estimates suggest that this increase was real; in 2009, the interval
was $[2069,2398]$ while in 2011, the entire interval shifted up to
$[2570,3068]$. In the case of Durham, we can confidently conclude
that residential burglaries increased -- why that increase occurred,
of course, is a different question.

\section{Conclusions}
A good deal of contemporary discussion about local crime
patterns in the U.S.\ is marred by unwarranted certitude
about the numbers and rates underlying that discussion.
Criminal justice officials, journalists, and even academic  
criminologists count crimes known to the police while 
ignoring key sources of uncertainty about those numbers.
Since the late 1960's and early 1970's, for example, it 
has been common criminological knowledge that many crimes 
are not reported to the police but somehow that knowledge 
ends up playing only a tangential role (if any role at 
all) in our public discourse about crime patterns at the 
local level.

Part of the problem is that there has been little 
methodological attention to the task of expressing and
transmitting uncertainty about crime patterns to policy
and lay audiences (Manski, 2003:21).
Based on Manski's work on bounds and
partial identification, however, we think it will be useful
for criminologists to begin reporting crime patterns in
terms of a range of uncertainty that expresses both what
is known and unknown about the numbers that are used to
measure those patterns. A key feature of the methods used 
here is that they explicitly abandon the goal of obtaining
point estimates in favor of a more realistic and reasonable
goal of obtaining interval estimates. Our approach 
provides one path by which criminologists can \emph{begin}
to reasonably express both what is known and unknown with 
current publicly available datasets. 

Another feature of our approach is that we move away from
the ``incredible certitude'' problem described by Charles
Manski (2011) -- the practice of developing unqualified 
and unjustifiably ``certain'' inferences based on weak data.
Criminologists are often asked by the media to comment on
small year-to-year movements in police-based crime statistics.
In our conversations with other criminologists, we have
noted that many feel quite uncomfortable characterizing this
or that small movement in the crime rate. The analysis in this
paper illustrates why these feelings of apprehension are justified.
As Eck and Riccio (1979) observed over 30 years ago, a movement 
of a few percentage points in the police statistics may or may 
not reflect real changes in crime. We think our approach to
this problem is useful because it allows us to transmit our 
uncertainty -- especially to lay audiences -- in systematic 
ways that have not been obvious in the past.

Still, there are limitations. First and foremost, we believe 
our bounds on the probability that a residential
burglary is reported to the police ($p_r$) are a reasonable 
starting point but improving our understanding of this 
interval would be constructive. This highlights an important 
direction for future research: achieving disciplinary consensus 
on the likely bounds for crime reporting probabilities should be 
a high priority. One reviewer of a previous version of this 
manuscript criticized our reporting rate intervals as being too
narrow.  That reviewer found it inconceivable that the local and 
national estimates would exhibit any particular comparability.
Most of the data that can be used to check on this
were collected in the 1970's in a series of city crime surveys
conducted by the National Criminal Justice Information and
Statistics Service (1974, 1975, 1976; see also Levitt, 1998)
and a research report by Lauritsen and Schaum (2005). While
there is not much local data to go on, it appears from the weight
of this evidence that most cities have residential burglary
reporting rates that are within a reasonably proximate range of
the national estimates of their time. The reviewer's comment 
nonetheless highlights the need for greater understanding of 
how closely local reporting rates track what is observed 
nationally. And -- if the reviewer turns out to be correct -- 
then the burglary rate intervals estimated in our work will 
be too narrow; a result which amplifies rather than diminishes 
our arguments.

We have considered several examples where point estimates based
on conventional methods prove to be highly misleading. Using those
methods one would draw the conclusion that one city had a higher
rate than another city or that a city's rate changed in a meaningful
way from one year compared to another. Our analysis shows that in
some of these comparisons, a plausible rival hypothesis cannot
be excluded: it is possible that the burglary rates are the 
same while only the reporting rate differs. Since the 
reporting rate, $p_r$, is not identified -- we can only make 
assumptions about its value -- the data cannot be used to 
resolve this ambiguity. Only information that reduces our 
uncertainty about the rate at which residential burglaries are 
reported to the police in the two cities will resolve it. 

The good news is that the development of this kind of 
information is feasible. The National Research Council along
with the BJS has recently considered 
a range of possibilities for improving on the small-area 
estimation capabilities of the NCVS (Groves and Cork, 2008;
Bureau of Justice Statistics, 2010). Most of the attention
has focused on small-area estimation of victimization rates
but further refinement of reporting rate estimates should
also be a priority. This is not a new idea -- Eck and Riccio
(1979) emphasized the possibilities of this approach decades
ago -- yet combining this emphasis with a focus on interval
estimation of crime rates may prove to be a viable way forward.
A key benefit of this kind of information would be a 
substantial reduction of the uncertainty  that is evident in 
our Figures 1-2.

It it noteworthy that we are able to make useful statements 
about residential burglary rates for North Carolina cities because 
the state reporting program clearly identifies residential burglaries 
known to the police. This is not done in the FBI's Uniform Crime Report
which presents counts of all burglaries -- both residential and
commercial -- known to the police in a single number. And we 
encounter difficulty using the FBI's burglary numbers since the 
NCVS only measures reporting behaviors for residential burglaries.
Expansion of our approach to other crimes will
require careful consideration of how the crimes described in the
UCR relate to the victimization incidents counted in the NCVS.
There is a well-developed literature on this topic (see, for 
example, Blumstein et al., 1991, 1992; 
Lynch and Addington, 2007) but there will be some 
difficulties in ensuring that the reporting probability gleaned
from the NCVS maps onto UCR crime categories in a meaningful way.
In our view, the field will be well served by taking on these
challenges.

We encountered a few other ambiguities in addition to the
reporting probability; namely, uncertainty due to the UCR's Hierarchy
Rule, the size of the population and the question of whether to
scale by the number of households or by the number of person 
(Gibbs and Erickson, 1976). 
It is surprising how large some of 
the differences in population estimates were and this uncertainty 
should be considered in more detail. We verified that each of the
jurisdictions we studied participated in the state Uniform Crime 
Reporting Program each month of each year during the 2009-2011 
calendar years (but Greensboro did not participate in the federal
program in 2011); still we clearly have no way to verify the 
accuracy of the numbers reported by the police departments 
(Westat, 2010:VII-VIII). This issue is always a threat to analyses
that rely on police-based crime statistics and our study is no
exception.

In our view, it will be useful for criminologists to: 
(1) be aware of the kinds of uncertainties discussed in 
this paper; (2) develop better information about uncertain 
parameters -- such as the probability of victimizations 
being reported to the police at the local level; 
(3) create analytic methods that will 
formally incorporate and transmit key sources of uncertainty 
in the measurement of crime rates; and (4) explore ways of 
conducting sensitivity analysis to assess the fragility of our
results. A fifth priority should be a program of research to consider
how identification problems such as those discussed in this paper
can be addressed within the framework of statistical models commonly
used to estimate effects of social and economic changes on crime rates.
Logically, there is no difference between a comparison of 
burglary rates in Charlotte and Raleigh from 2009 to 2010 and the kinds
of panel regresssion, difference-in-difference, and 
pooled-cross-sectional time series estimators commonly used to 
identify causal relationships in crime data. All of the uncertainties
discussed here are present in space-and-time crime regressions
commonly estimated by criminologists. Yet the issues discussed in
this paper loom as major sources of uncertainty for these models.
We view our approach as an initial, constructive, and necessary 
step in the direction of a more balanced and informative use of 
aggregate crime statistics.

\newpage
\section*{References}
\normalsize
\begin{trivlist}
\itemsep=0.15in
\item
Addington, Lynn A.\ (2007).\
Using NIBRS to study methodological sources of divergence
between the UCR and NCVS. In \emph{Understanding crime statistics:
revisiting the divergence of the NCVS and UCR}, pp.\ 225-250.
New York: Cambridge University Press.

\item
American Society of Criminology (2007).\
\href{http://www.asc41.com/policies/policyPositions.html}
{Official policy position of the Executive Board of the American 
Society of Criminology with respect to the use of Uniform 
Crime Reports data.}
Columbus, OH: American Society of Criminology.

\item
Anderson, Elijah (1999). \
\emph{Code of the street: Decency, violence, and the moral 
life of the inner city.} New York: W.W. Norton.

\item
Baumer, Eric P.\ and Janet L.\ Lauritsen (2010).
Reporting crime to the police 1973-2005: A multivariate analysis of 
long-term trends in the National Crime Survey (NCS) and National
Crime Victimization Survey (NCVS). 
\emph{Criminology}, 48:131-185.

\item
Biderman, Albert D.\ and Albert J.\ Reiss Jr.\ (1967). On
exploring the ``dark figure'' of crime. \emph{Annals of the
American Academy of Political and Social Science}, 374:733-748.

\item
Blumstein, Alfred, Jacqueline Cohen, and Richard Rosenfeld (1991).\
Trend and deviation in crime rates: a comparison of UCR and
NCS data for burglary and robbery. \emph{Criminology}, 29:237-263.

\item
Blumstein, Alfred, Jacqueline Cohen, and Richard Rosenfeld (1992).\
The UCR-NCS relationship revisited: a reply to Menard.
\emph{Criminology}, 30:115-124.

\item
Blumstein, Alfred and Richard Rosenfeld (2008).
Factors contributing to U.S.\ crime trends.
In \emph{Understanding crime trends}, pp.\ 13-44.
Washington, DC: National Academy Press.

\item
Brier, Stephen S.\ and Stephen E.\ Fienberg (1980).
Recent econometric modeling of crime and punishment:
support for the deterrence hypothesis. \emph{Evaluation
Review}, 4:147-191.

\item
Bureau of Justice Statistics (2010). 
\emph{Questions and answers from the 4/27/10 NCVS redesign 
session on sub-national estimates}. 
Washington, DC: U.S. Department of Justice.

\item
Bialik, Carl (2010). In crime lists, nuance is a victim. 
\emph{Wall Street Journal} (published December 4, 2010).

\item
Carbon, Susan B.\ (2012). Statement before the Committee 
on the Judiciary of the United States House of 
Representatives of February 16, 2012.\ Washington, DC: 
U.S.\ Department of Justice.

\item
Catalano, Shannan M.\ (2004). \emph{Criminal victimization, 2003.}
Washington, DC: U.S.\ Department of Justice.

\item
Catalano, Shannan M.\ (2005). \emph{Criminal victimization, 2004.}
Washington, DC: U.S.\ Department of Justice.

\item
Catalano, Shannan M.\ (2006). \emph{Criminal victimization, 2005.}
Washington, DC: U.S.\ Department of Justice.

\item
Eck, John E.\ and Lucius J.\ Riccio (1979). Relationship between 
reported crime rates and victimization survey results: an
empirical and analytical study. \emph{Journal of Criminal
Justice}, 7:293-308.

\item
Fagan, Jeffrey and Tom R. Tyler.\ (2005).\
Legal socialization of children and adolescents. 
\emph{Social  Justice Research} 18:217-242.
   
\item
Federal Bureau of Investigation (2009). 
\emph{The nation's two crime measures.}
Washington, DC: U.S.\ Department of Justice.

\item
Gibbs, Jack P.\ and Maynard L.\ Erickson (1976).
Crime rates of American cities in an ecological context.
\emph{American Journal of Sociology}, 82:605-620.

\item
Groves, Robert M.\ and Daniel L.\ Cork (eds.) (2008).
\emph{Surveying victims: options for conducting the i
National Crime Victimization Survey.}
Washington, DC: National Academy Press.

\item
James, Nathan and Logan Rishard Council (2008).
\emph{How crime in the United States is measured.}
Washington, DC: Congressional Research Service.

\newpage
\item
Kirk, David S. and Mauri Matsuda (2011).
Legal cynicism, collective efficacy, and the ecology of arrest.
\emph{Criminology}, 49:443-472.

\item
Kirk, David S. and Andrew V.\ Papachristos (2011).
Cultural mechanisms and the persistence of neighborhood violence. 
\emph{American Journal of Sociology} 116: 1190-1233.

\item
Lab, Steven P., Marian R.\ Williams, Jefferson E.\ Holcomb,
Melissa W.\ Burek, William R.\ King, and Michael E.\ Buerger
(2008). \emph{Criminal justice: the essentials}. New York:
Oxford University Press.
   
\item
Lauritsen, Janet L.\ and Robin J.\ Schaum (2005).
\emph{Crime and victimization in the three largest
metropolitan areas, 1980-1998.}
Washington, DC: U.S. Department of Justice.

\item
Levitt, Steven D.\ (1998). The relationship between
crime reporting and police: implications for the use
of Uniform Crime Reports. \emph{Journal of Quantitative
Criminology}, 14:61-81.

\item
Lohr, Sharon and N.G.N.\ Prasad (2003).
Small area estimation with auxiliary survey data.
\emph{Canadian Journal of Statistics}, 31:383-396.

\item
Lynch, James P.\ and Lynn A.\ Addington (eds.) (2007).
\emph{Understanding crime statistics: revisiting the divergence
of the NCVS and UCR.} New York: Cambridge University Press.

\item
Lyttle, Steve (2012).
CMPD says crime drops 7.1\% in 2011. 
\emph{The Charlotte Observer}, Wednesday January 18, 2012.

\item
Maltz, Michael D.\ (1975). Crime statistics: a mathematical
perspective. \emph{Journal of Criminal Justice}, 3:177-194.

\item
Maltz, Michael D.\ (1999).
\emph{Bridging gaps in police crime data.}
Washington, DC: U.S.\ Department of Justice.

\item
Maltz, Michael D.\ and Joseph Targonski (2002).
Measurement and other errors in county-level UCR data: 
a reply to Lott and Whitley.
\emph{Journal of Quantitative Criminology}, 19:199-206.

\newpage
\item
Manski, Charles F.\ (1995).
\emph{Identification problems in the social sciences.}
Cambridge, MA: Harvard University Press.

\item
Manski, Charles F.\ (2003).
\emph{Partial identification of probability distributions.}
New York: Springer-Verlag.

\item
Manski, Charles F.\ (2011).
Policy analysis with incredible certitude.
\emph{The Economic Journal}, 121:F261-F289.

\item
Merton, Robert K.\ (1968).
\emph{ Social Theory and Social Structure}. 
New York: Free Press.
   
\item 
National Criminal Justice Information and Statistics Service (1974).
\emph{Criminal victimization surveys in the nation's five largest cities.}
Washington, DC: U.S.\ Department of Justice.

\item 
National Criminal Justice Information and Statistics Service (1975).
\emph{Criminal victimization surveys in 13 American cities.}
Washington, DC: U.S.\ Department of Justice.

\item 
National Criminal Justice Information and Statistics Service (1976).
\emph{Criminal victimization surveys in 8 American cities.}
Washington, DC: U.S.\ Department of Justice.

\item
Nelson, James F.\ (1979). Implications for the ecological study of
crime: a research note. In William H.\ Parsonage (ed.), \emph{Perspectives
on Victimology}, pp.\ 21-28. Beverly Hills, CA: Sage Publications.

\item
Papachristos, Andrew V., Tracey Meares and Jeffrey Fagan (2011).
Why do offenders obey the law? The influence of legitimacy 
and social networks on active gun offenders. Unpublished paper.

\item
Paternoster, Raymond, Robert Brame, Ronet Bachman, 
and Lawrence W.\ Sherman (1997). 
Do fair procedures matter? The effect of procedural 
justice on spouse assault. 
\emph{Law and Society Review}, 31:163-204.
  
\item
Pepper, John V.\ (2001). How do response problems affect 
survey measurement of trends in drug use? In 
\emph{Informing America's Policy on Illegal Drugs: What 
We Don't Know Keeps Hurting Us}, pp. 321-347. 
Washington, DC: National Academy Press.

\item
Pepper, John V.\ and Carol V.\ Petrie (eds.) (2003).
\emph{Measurement problems in criminal justice research:
workshop summary}.
Washington, DC: National Academy Press.

\item
Raghunathan, Trivellore E., Dawei Xie, 
Nathaniel Schenker, Van L. Parsons, William W. Davis, 
Kevin W. Dodd and Eric J. Feuer (2007).
Combining information from two surveys to
estimate county-level prevalence rates of cancer
risk factors and screening.
\emph{Journal of the American Statistical Association},
102:474-486.

\item
Rand, Michael R.\ and Callie M.\ Rennison (2002).
True crime stories? Accounting for differences in our national
crime indicators.
\emph{Chance}, 15:47-51.

\item
Rantala, Ramona R.\ (2000). 
Effect of NIBRS on crime statistics.
Washington, DC: U.S.\ Department of Justice.

\item
Reisig, Michael D., Scott E. Wolfe and Kristy Holtfreter (2011).
Legal cynicism, legitimacy,  and criminal offending.
\emph{Criminal Justice and Behavior}, 38:1265-1279.

\item
Rennison, Callie M.\ and Michael R.\ Rand (2003).
\emph{Criminal victimization, 2002.}
Washington, DC: U.S.\ Department of Justice.

\item
Rosenfeld, Richard and Janet L.\ Lauritsen (2008).
The most dangerous crime rankings. 
\emph{Contexts}, 7:66-67.

\item
Sampson, Robert J.\ and Dawn Jeglum Baertusch. (1998).
Legal cynicism and (subcultural?) tolerance of deviance: 
The neighborhood context of racial differences.
\emph{Law and Society Review}, 32:777-804.

\item
Skogan, Wesley G.\ (1974).\ The validity of official
crime statistics: An empirical investigation.
\emph{Social Science Quarterly}, 55:25-38.

\item
Slocum, Lee Ann, Terrance J. Taylor, Bradley T. Brick, and
Finn-Aage Esbensen. (2010). 
Neighborhood structural characteristics, 
individual-level attitudes, and youths crime reporting intentions. 
\emph{Criminology}, 48:1063-1100.
  
\item
State Bureau of Investigation (North Carolina). (2012).
\emph{Crime in North Carolina, 2010).} 
Raleigh, NC: North Carolina Department of Justice.

\item
Sunshine, Jason and Tom R. Tyler.\ (2003). 
The role of procedural justice and legitimacy in 
shaping public support for policing. 
\emph{Law and Society Review} 37: 513-547.
 
\item
Swidler, Ann (1986). 
Culture in action: Symbols and strategies. 
\emph{American Sociological Review}, 51:273-286.
   
\item
Truman, Jennifer L.\ and Michael R.\ Rand (2010).
\emph{Criminal victimization, 2009.}
Washington, DC: U.S.\ Department of Justice.

\item
Truman, Jennifer L.\ (2011).
\emph{Criminal victimization, 2010.}
Washington, DC: U.S.\ Department of Justice.

\item
Truman, Jennifer L.\ and Michael Planty (2012).
\emph{Criminal victimization, 2011.}
Washington, DC: U.S.\ Department of Justice.

\item
Tyler, Tom R.\ (2006). 
\emph{Why people obey the law}. 
New Haven, CT: Yale University Press.
   
\item
Tyler, Tom R.\ and Jeffrey Fagan. (2008). 
Legitimacy and cooperation: 
Why do people help the police fight crime in their communities? 
\emph{Ohio State Journal of Criminal Law}, 6:231-274.

\item
U.S.\ Census Bureau (2012). \emph{State and County QuickFacts.}
Washington, DC: U.S.\ Department of Commerce.

\item
Wainer, Howard (1986). The SAT as a social indicator: a pretty
bad idea. In Howard Wainer (ed.), \emph{Drawing Inferences 
from Self-Selected Samples}, pp. 7-21. New York: Springer-Verlag.

\item
Westat (2010).\ \emph{NCVS task 4 report: summary of
options relating to local area estimation.}
Rockville, MD: Westat.
  
\end{trivlist}
\end{document}